\documentclass[11pt,epsf,letterpaper]{article}%
\usepackage[usenames,dvipsnames,svgnames,table]{xcolor}
\usepackage{color}
\usepackage{amsmath}
\usepackage{amsfonts}
\usepackage{verbatim}
\usepackage{amssymb}
\usepackage{graphicx}
\usepackage{epstopdf}
\usepackage{mathrsfs}
\usepackage{xcolor}
\usepackage{cite}
\usepackage{subcaption}%
\providecommand{\U}[1]{\protect\rule{.1in}{.1in}}

\begin{document}

\title{Universal Formula for the Holographic Speed of Sound}
\author{$^{(1,2)}$Andr\'{e}s Anabal\'{o}n, $^{(3)}$Tom\'{a}s Andrade, $^{(4,5)}%
$Dumitru Astefanesei and $^{(6,7)}$Robert Mann\\\textit{$^{(1)}$Departamento de Ciencias, Facultad de Artes Liberales,}\\\textit{Universidad Adolfo Ib\'{a}\~{n}ez, Vi\~{n}a del Mar, Chile.} \\\textit{$^{(2)}$DISAT, Politecnico di Torino, Corso Duca degli Abruzzi 24,
I-10129 Torino, Italia.} \\\textit{$^{(3)}$Rudolf Peierls Center for Theoretical Physics, University of
Oxford,}\\\textit{1 Keble Road, Oxford OX1 3NP, UK} \\\textit{$^{(4)}$Instituto de F\'\i sica, Pontificia Universidad Cat\'olica de
Valpara\'\i so,} \\\textit{ Casilla 4059, Valpara\'{\i}so, Chile.}\\\textit{$^{(5)}$Max-Planck-Institut f\"ur Gravitationsphysik,
Albert-Einstein-Institut, 14476 Golm, Germany}\\\textit{$^{(6)}$ Department of Physics and Astronomy, University of Waterloo,}\\\textit{Waterloo, Ontario, Canada N2L 3G1.}\\\textit{$^{(7)}$ Perimeter Institute, 31 Caroline Street North Waterloo,
Ontario Canada N2L 2Y5.}}
\maketitle

\begin{abstract}
We consider planar hairy black holes in five dimensions with a real scalar
field in the Breitenlohner-Freedman window and show that is possible to derive
a universal formula for the holographic speed of sound for any mixed boundary
conditions of the scalar field. As an example, we locally construct the most
general class of planar black holes coupled to a single scalar field in the
consistent truncation of type IIB supergravity that preserves the $SO(3)\times
SO(3)$ R-symmetry group of the gauge theory. We obtain the speed of sound for
different values of the vacuum expectation value of a single trace operator
when a double trace deformation is induced in the dual gauge theory. In this
particular family of solutions, we find that the speed of sound exceeds the
conformal value. Finally, we generalize the formula of the speed of sound to
arbitrary dimensional scalar-metric theories whose parameters lie within the
Breitenlohner-Freedman window.

\end{abstract}

\section{Introduction}

An important achievement of AdS/CFT duality \cite{Maldacena:1997re} is the
holographic computation of the universal value of the ratio of the shear
viscosity to entropy density \cite{Policastro:2001yc}. Finite 't~Hooft
coupling corrections on the gauge theory side correspond to $\alpha^{\prime}$
corrections on the gravity side, inclusion of which will modify the ratio thus
violating the conjectured viscosity bound \cite{Kats:2007mq, Brigante:2007nu}.
That seems to indicate a strong link between the actual form of the Lagrangian
and the holographic transport coefficients and this relation was implicit in
the further exploration of the field. The computation of the speed of sound
\cite{Gursoy:2007cb, Gursoy:2007er, Gubser:2008ny, Gubser:2008yx, He:2011hw}
in the dual theory assumes this idea and a bound was also supposed to exist
\cite{Cherman:2009tw}. The fact that different gravity Lagrangians are the key
to describing different speeds of sound was extensively used in the literature
until recently (see for instance \cite{Hoyos:2016cob}).

The main goal of this letter is to obtain a universal formula for the the
holographic speed of sound. The crucial point is that when the mass of the
scalar field is in the Breitenlohner-Freedman (\textbf{BF}) window
\cite{Breitenlohner:1982bm, BF}
\begin{equation}
-\frac{(D-1)^{2}}{4l^{2}}=m_{BF}^{2}\leq m^{2}<m_{BF}^{2}+l^{-2}\text{ ,}
\label{window}%
\end{equation}
the holographic speed of sound depends as much on the details of the scalar
field potential as it does on the boundary conditions that the scalar field
satisfies. The universal formula we obtain takes the details of this relation
into account.

In order to do this, we face the problem of solving the bulk theory. We tackle
this by means of a generalization of the designer gravity soliton line
\cite{Hertog:2004ns}, an idea already considered in \cite{Liu:2013gja} and
also discussed in \cite{Anabalon:2016yfg}. The key point is that scalar fields
satisfying (\ref{window}) admit an infinite number of possible boundary
conditions, which can be traced back to the existence, in any dimension, of
two normalizable modes \cite{Ishibashi:2004wx}. In general, these boundary
conditions can break the conformal symmetry in the boundary
\cite{Henneaux:2006hk} (the implications of these boundary conditions for the
energy of the hairy black holes and their thermodynamics were investigated in
great detail in \cite{Amsel:2006uf, Anabalon:2014fla}). When the scalar field
mass is in the BF window (\ref{window}), it provides two asymptotic
integration constants ($\alpha,\beta$) to the system. In addition, a static
black hole metric provides one extra integration constant, $\mu$, related to
its mass. The solution space of the fully back-reacted metric plus the scalar
field is characterized by three integration constants \cite{Liu:2015tqa}. The
solution of the non-linear system of differential equations provides the map%
\begin{equation}
\alpha=\alpha(\varphi_{h},\mathcal{A})\text{,\qquad}\beta=\beta(\varphi
_{h},\mathcal{A})\text{,\qquad}\mu=\mu(\varphi_{h},\mathcal{A})\text{,}
\label{map}%
\end{equation}
where $\left(  \varphi_{h},\mathcal{A}\right)  $ are the horizon data, namely
the value of the scalar field at the horizon, $\varphi_{h}$, and the
normalized black hole area\footnote{We shall consider planar horizons with one
compact direction. In this case $\frac{\mathcal{A}}{4G}$ is the entropy
density.}, $\mathcal{A}$. The proof that dependence on $\mathcal{A}$ of the
map \eqref{map} is universal and related only to the mass of the scalar field
is the second essential result of this letter.

In this article we focus on finite temperature solutions. For the zero
temperature case, a similar analysis could be done in the lines of
\cite{Anabalon:2013sra}, with the introduction of an $AdS_{2}$ factor in the
near horizon geometry \cite{Sen:2005wa, Sen:2007qy, Astefanesei:2006dd}.

We begin with a concise review of deformations of the gauge theory along the
lines of \cite{Witten:2001ua}. We explicitly construct the map \eqref{map}
that allows us to obtain a black hole line that can be used to get a general
formula for the speed of sound. This formula is universal as is valid for all
boundary conditions. Then, we present an explicit example in type IIB
supergravity and discuss the physics of a marginal deformation. We discuss how
our result is valid for any value of the coupling constant of the deformation
and its relation to the renormalization group (\textbf{RG}) flow in the gauge
theory. It is worth remarking that this is different than the usual RG flow of
the relativistic holographic fluid and speed of sound at a fixed cut-off which
were presented in \cite{Kuperstein:2011fn, Kuperstein:2013hqa}. In the last
section we provide a formula for any scalar field with a mass in the range (1).

\section{The Gauge Theory Deformation and Holography}

Here we briefly review the relation between the field theory deformation and
its holographic interpretation. In field theory one would like to add to the
action a functional of the form \cite{Witten:2001ua}%

\begin{equation}
\Delta I=\left(  \frac{N}{2\pi}\right)  ^{2}\int_{\partial M}W(\mathcal{O}%
)\sqrt{-\gamma^{\left(  0\right)  }}d^{4}x\text{,}%
\end{equation}
where $W(\mathcal{O})$ is an arbitrary function of a single trace operator
$\mathcal{O}$. Gravitational variables are recovered with the standard AdS/CFT
identification $N^{2}=\frac{\pi}{2}\frac{l^{3}}{G}$ \cite{Maldacena:1997re}.
The bulk metric is%

\begin{equation}
ds^{2}=g_{\mu\nu}dx^{\mu}dx^{\nu}=\frac{l^{2}}{r^{2}}dr^{2}+\frac{r^{2}}%
{l^{2}}\gamma_{ab}dx^{a}dx^{b}\,,
\end{equation}
where $\gamma_{ab}=\gamma_{ab}^{\left(  0\right)  }+O(r^{-2})$. The
deformation in five dimensional variables yields%

\begin{equation}
\Delta I=\frac{l^{3}}{\kappa}\lim_{r\rightarrow\infty}\int_{\partial
M}W(\mathcal{O})\sqrt{-\gamma}d^{4}x\text{ ,}%
\end{equation}
where $\kappa=8\pi G$ is the reduced Newton constant in five dimensions. The
gravitational action is%

\begin{equation}
\kappa I\left[  g,\varphi\right]  =\int_{M}d^{5}x\sqrt{-g}\left[  \frac{R}%
{2}-\frac{1}{2}\left(  \partial\varphi\right)  ^{2}-V(\varphi)\right]
+I_{ct}+I_{ct}^{\varphi}+\kappa\Delta I\text{ ,}%
\end{equation}
where $I_{ct}$ and $I_{ct}^{\varphi}$ are counterterms that render the
gravitational action finite and the variational principle well posed
\cite{Anabalon:2016yfg, Henningson:1998gx, Balasubramanian:1999re,
Mann:1999pc, Bianchi:2001de, Bianchi:2001kw, Marolf:2006nd,
Papadimitriou:2007sj, Anabalon:2015xvl}. When the mass of the scalar field
saturates the BF bound its fall-off is
\begin{equation}
\varphi=\frac{\alpha l^{4}}{r^{2}}\ln\left(  \frac{r}{r_{0}}\right)
+\frac{\beta l^{4}}{r^{2}}+O\left(  \frac{\ln\left(  r\right)  ^{2}}{r^{3}%
}\right)  \label{phi1}%
\end{equation}
where $\alpha$ and $\beta$ are normalized to have engineering dimension $2$.
The relevant scalar counterterm is \cite{Anabalon:2016yfg, Marolf:2006nd}%

\begin{equation}
I_{ct}^{\varphi}=\lim_{r\rightarrow\infty}\left[  \int_{\partial M}\frac{1}%
{2}\varphi n^{\mu}\partial_{\mu}\varphi\sqrt{-h}d^{4}x-l^{3}\int_{\partial
M}\frac{\alpha\beta}{2}\sqrt{-\gamma}d^{4}x\right]
\end{equation}
where $h_{\mu\nu}=\frac{r^{2}}{l^{2}}\gamma_{ab}$ and $n$ is the outward
pointing normal to $h$. In this case the scalar contribution to the
Euler-Lagrange variation is%
\begin{equation}
\delta I=\frac{l^{3}}{\kappa}\lim_{r\rightarrow\infty}\int_{\partial M}\left(
\frac{dW(\beta)}{d\beta}-\alpha\right)  \delta\beta\sqrt{-\gamma}d^{4}x
\end{equation}
where we have identified the vacuum expectation value (\textbf{VEV}) of the
single trace operator with $\beta$ ($\left\langle \mathcal{O}\right\rangle
=\beta$). The source associated with the VEV is $J=\alpha-\frac{dW(\beta
)}{d\beta}$ \cite{Marolf:2006nd}. This follows from the transformed action%

\begin{equation}
\bar{I}=I+\frac{l^{3}}{\kappa}\lim_{r\rightarrow\infty}\int_{\partial
M}J\mathcal{O}\sqrt{-\gamma}d^{4}x\text{ .}%
\end{equation}
In this case the variation of the partition function at $J=0$ yields the result%

\begin{equation}
\frac{\kappa}{l^{3}}\left.  \frac{1}{Z(J)}\frac{\delta Z(J)}{\delta
J}\right\vert _{J=0}=\left\langle \mathcal{O}\right\rangle =\beta\text{ .}%
\end{equation}
When the field theory is at finite temperature, different deformations
correspond to different thermodynamic properties that are constrained by the
bulk dynamics. We shall analyze this in the next section.

\section{The black hole surface}

In this section we shall outline the construction of the surface defined by
the embedding (\ref{map}). We consider the class of metrics in five
dimensions
\begin{equation}
ds^{2}=e^{A}(-fdt^{2}+d\Sigma)+e^{B}\frac{dr^{2}}{f}\text{ ,} \label{metclass}%
\end{equation}
where $d\Sigma$ is a Ricci flat surface. For static metrics is possible to
introduce the new variables $Z=\frac{d\varphi}{dA}~$and $Y=Zf\frac{d\varphi
}{df}$ and show that the Einstein equations are satisfied if
\begin{equation}
\frac{dZ}{d\varphi}=\left(  3\frac{dV}{d\varphi}+4ZV\right)  \frac{\left(
2Z^{2}Y-6Y-3Z^{2}\right)  }{12VZY}\text{ ,}\qquad\frac{dY}{d\varphi}=\left(
3\frac{dV}{d\varphi}+2ZV\right)  \frac{\left(  2Z^{2}Y-6Y-3Z^{2}\right)
}{6VZ^{2}}\text{ .}%
\end{equation}
It follows that $\frac{dZ}{d\varphi}$ is finite at the horizon, located at
$Y(\varphi_{h})=0$, if and only if $Z\left(  \varphi_{h}\right)  =-\left.
\frac{3}{4V}\frac{dV}{d\varphi}\right\vert _{\varphi=\varphi_{h}}$.
\ Furthermore, we readily see that these equations decouple and reduce to the
single master equation%
\begin{align}
-3Z\left(  3\frac{dV}{d\varphi} + 4ZV\right)  \frac{d^{2}Z}{d\varphi^{2}}  &
+\left(  -9\frac{dV}{d\varphi} +12ZV\right)  \left(  \frac{dZ}{d\varphi
}\right)  ^{2}\nonumber\\
&  +\left[  \left(  Z^{2}+3\right)  8ZV+\left(  18Z^{2}+1\right)  \frac
{dV}{d\varphi}+9Z\frac{d^{2}V}{d\varphi^{2}}\right]  \frac{dZ}{d\varphi
}=0\text{ .}%
\end{align}
The regularity conditions imply that $Z$ is completely determined by the value
of $\varphi$ at the horizon, and so
\begin{equation}
A\left(  r\right)  =\int_{\varphi_{h}}^{\varphi(r)}\frac{d\varphi}{Z}+\frac
{2}{3}\ln\left(  \mathcal{A}\right)  \text{ ,}%
\end{equation}
which follows from the definition of $Z$ and the value of $A(r_{h})$. It is a
consequence of the fall-off of the scalar (\ref{phi1}) that, asymptotically,
$e^{A}\varphi\sim\alpha l^{2}\ln(\frac{r}{r_{0}}).$ The derivative of this
relation with respect to $\ln r\sim\frac{A}{2}$ yields $2e^{A}\left(
\varphi+\frac{d\varphi}{dA}\right)  \sim\alpha l^{2}$. Thus,
\begin{equation}
l^{2}\alpha\left(  \varphi_{h},\mathcal{A}\right)  ={\lim_{\varphi
\rightarrow0}}2\left(  Z+\varphi\right)  e^{A}=\alpha_{h}\left(  \varphi
_{h}\right)  \mathcal{A}^{\frac{2}{3}}\text{ .} \label{map 1}%
\end{equation}
We see that $\alpha$ is generically a function of $\varphi_{h}$ times a very
precise function of the normalized black hole area. When there exists a black
hole the metric fall-off can be taken as%

\begin{equation}
e^{A}f=\frac{r^{2}}{l^{2}}-\frac{\mu l^{6}}{r^{2}}+O(r^{-3})\text{ ,}%
\qquad\text{ }e^{A}=\frac{r^{2}}{l^{2}}+O(r^{-3})\text{ ,}\qquad\text{ }%
\frac{e^{B}}{f}=\frac{l^{2}}{r^{2}}+O(\frac{\ln\left(  r\right)  ^{2}}{r^{6}})
\label{expa}%
\end{equation}
where the normalization of the integration constant $\mu$ is chosen to match
its mass scaling dimension, $\Delta\left(  \mu\right)  =4$ with its
engineering dimension. In this gauge, the introduction of the scalar field
yields a dual perfect fluid with energy momentum tensor
\cite{Anabalon:2016yfg}
\begin{equation}
-\frac{2}{\sqrt{-\gamma^{\left(  0\right)  }}}\frac{\delta I}{\delta
\gamma^{ab(0)}}=\left\langle \mathcal{T}_{ab}\right\rangle =\left(
\rho+p\right)  u_{a}u_{b}+p\gamma_{ab}^{(0)},
\end{equation}
where
\begin{equation}
p=\frac{l^{3}}{\kappa}\left[  \frac{\mu}{2}+\frac{1}{8}\left(  \alpha
^{2}-4\alpha\beta+8W(\beta)\right)  \right]  \label{press}%
\end{equation}%
\begin{equation}
\rho=\frac{l^{3}}{\kappa}\left[  \frac{3\mu}{2}-\frac{1}{8}\left(  \alpha
^{2}-4\alpha\beta+8W(\beta)\right)  \right]  \label{dens}%
\end{equation}
with $u=\partial_{t}$. Note that when the scalar field vanishes (\ref{dens})
we recover the standard relation $\rho=3p$ for a thermal gas of massless
particles \cite{Myers:1999psa}. The temperature of the configuration is given
by%
\begin{equation}
T=\frac{\mathcal{A}^{\frac{1}{3}}e^{\frac{1}{2}B(r_{h})}}{2\pi}\left\vert
\frac{dV\left(  \varphi_{h}\right)  }{d\varphi}\right\vert \text{ ,}%
\end{equation}
It follows from the equations of motion and the boundary conditions that
$B(r_{h})$ depends only on $\varphi_{h}$. The Smarr formula gives
\cite{Liu:2015tqa}%

\begin{equation}
\rho+p=Ts=\frac{2\mu l^{3}}{\kappa}\text{ .}%
\end{equation}
where $s=\frac{\mathcal{A}}{4G}$ . The second map is then%
\begin{equation}
\mu\left(  \varphi_{h},\mathcal{A}\right)  =l^{-4}\mu_{h}\left(  \varphi
_{h}\right)  \mathcal{A}^{\frac{4}{3}}\text{ .} \label{mu}%
\end{equation}
Inserting (\ref{map}) in the first law of black hole thermodynamics,
$\delta\rho=T\delta s$, with the knowledge of (\ref{mu}) and (\ref{map 1}), it
shows that the terms proportional to $\delta\mathcal{A}$ cancel, provided
\begin{equation}
\beta\left(  \varphi_{h},\mathcal{A}\right)  =-\frac{1}{2}\alpha\ln\left\vert
\frac{\alpha}{\alpha_{0}}\right\vert +\alpha z_{h}\left(  \varphi_{h}\right)
~, \label{beta}%
\end{equation}
where $z_{h}(\varphi_{h})$ is a function of the value of the scalar at the
horizon. The terms proportional to the variation of $\delta\varphi_{h}$ cancel
if and only if
\begin{equation}
-3d\mu_{h}+\alpha_{h}^{2}dz_{h}=0\text{ .} \label{mu0}%
\end{equation}
The equations (\ref{map 1}), (\ref{mu}) and (\ref{beta}) define the black hole
surface. Deformations that intersect the black hole surface can be associated
with thermal states when the field theory is at finite temperature.

\section{Universal formula for the speed of sound}

A useful consequence of the general considerations made so far, it is that
there are two $\mathcal{A}-$independent functions
\begin{equation}
z\equiv\frac{\beta}{\alpha}+\frac{1}{2}\ln\left\vert \frac{\alpha}{\alpha_{0}%
}\right\vert \text{ },\qquad\frac{\mu}{\alpha^{2}}\text{ }. \label{Aind}%
\end{equation}
The characterization of the speed of sound for a single scalar field theory
can then be reduced to finding the black hole line%
\begin{equation}
\lim_{r\rightarrow\infty}\frac{\left(  1-f\right)  A^{2}}{4\varphi^{2}}%
=\frac{\mu}{\alpha^{2}}=F(z)\,\text{.} \label{BHL}%
\end{equation}
AdS-invariant boundary conditions correspond to $z=z^{\ast}$, where $z^{\ast}$
is a fixed number. In this case the energy density, pressure and speed of
sound are automatically%
\begin{equation}
\rho=\frac{3l^{3}}{2\kappa}F(z^{\ast})\alpha^{2},\qquad p=\frac{l^{3}}%
{2\kappa}F(z^{\ast})\alpha^{2}\text{ },\qquad c_{s}^{2}=\left(  \frac{\partial
p}{\partial\rho}\right)  _{J}=\frac{1}{3}%
\end{equation}
where the partial derivative is taken at fixed source, $J=\alpha
-\frac{dW(\beta)}{d\beta}=0$. It follows that there are infinite number of
theories with $c_{s}^{2}=\frac{1}{3}$. To move away from this point one should
consider a generic boundary condition, of the form $z=\omega(\alpha)$, the
knowledge of the black hole line (\ref{BHL}) allows the construction of the
derivative of the pressure and density at fixed source and finally from
\eqref{press} and \eqref{dens} a formula for the speed of sound:%

\begin{equation}
\left(  \frac{\partial p}{\partial\alpha}\right)  _{J}=\frac{l^{3}}{\kappa
}\left[  \alpha F+\frac{\alpha^{2}\dot{F}\omega^{\prime}}{2}+\frac{1}{2}%
\alpha^{2}\omega^{\prime}\right]  \text{ },\qquad\left(  \frac{\partial\rho
}{\partial\alpha}\right)  _{J}=\frac{l^{3}}{\kappa}\left[  3\alpha F+\frac
{3}{2}\alpha^{2}\dot{F}\omega^{\prime}-\frac{1}{2}\alpha^{2}\omega^{\prime
}\right]
\end{equation}
\begin{equation}
c_{s}^{2}=\left(  \frac{\partial p}{\partial\rho}\right)  _{J}=\frac{1}%
{3}+\frac{4}{3}\frac{\alpha^{2}\omega^{\prime}}{6\alpha F+3\alpha^{2}\dot
{F}\omega^{\prime}-\alpha^{2}\omega^{\prime}} \label{speed1}%
\end{equation}
where $\dot{F}=\frac{dF}{dz},$ $\omega^{\prime}=\frac{d\omega}{d\alpha}.$ It
follows that the speed of sound can be fully characterized using the black
hole line defined by $F$ and the boundary condition defined by $\omega
(\alpha)$.

In holographic terms, (\ref{speed1}) yields the dependence of the speed of
sound on the VEV $\beta$ associated with the source $J$. Note that the formula
(\ref{speed1}) is universal and can be constructed for any theory in terms of
its black hole line and boundary condition. It is worth mentioning that the
second term in \eqref{speed1} does not have a definite sign, so a priori it is
not clear if the speed of sound for these theories will be bounded by the
conformal value $c_{s}^{2} = 1/3$. We also note that (\ref{speed1}) indeed
reproduces the result for a gas of massless particles for AdS invariant
boundary conditions where $\omega^{\prime}=0$.

In order to further study the thermodynamics for different boundary conditions
one needs to construct the black hole line. A minimal example that captures
the ideas discussed here is a consistent single scalar field truncation of
maximal supergravity in five dimensions:%

\begin{equation}
V\left(  \varphi\right)  =-\frac{3}{2l^{2}}\left[  3+\cosh\left(  \frac
{2\sqrt{6}}{3}\varphi\right)  \right]  \text{ ,} \label{V cosh}%
\end{equation}
which breaks the isometries of the $S^{5}$ to $SO(3)\times SO(3)$
\cite{Freedman:1999gk}. Using the standard AdS/CFT dictionary, the {operator
dual to $\beta$ is $\mathcal{O}$}%
\begin{equation}
\mathcal{O=}\frac{1}{N}\text{Tr}\left(  \phi_{1}^{2}+\phi_{2}^{2}+\phi_{3}%
^{2}-\phi_{4}^{2}-\phi_{5}^{2}-\phi_{6}^{2}\right)
\end{equation}
where the $\phi_{I}$ are the Super Yang-Mills scalars \cite{Witten:1998qj}.

\begin{figure}[th]
\centering
\begin{subfigure}{0.4\linewidth}
\includegraphics[width= \linewidth]{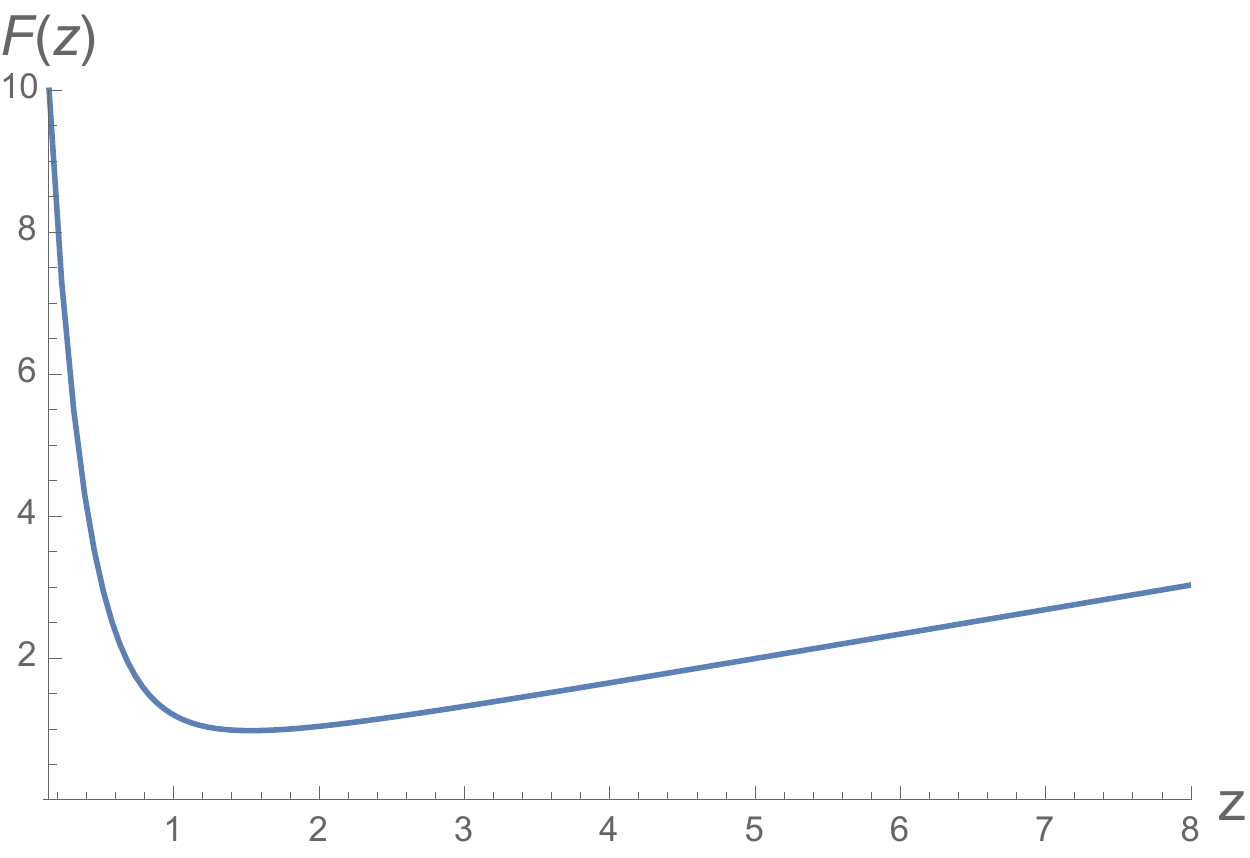}
\caption{\label{fig1} Black hole line defined by equation \eqref{BHL}.}
\end{subfigure} \qquad\qquad\begin{subfigure}{0.4\linewidth}
\includegraphics[width= \linewidth]{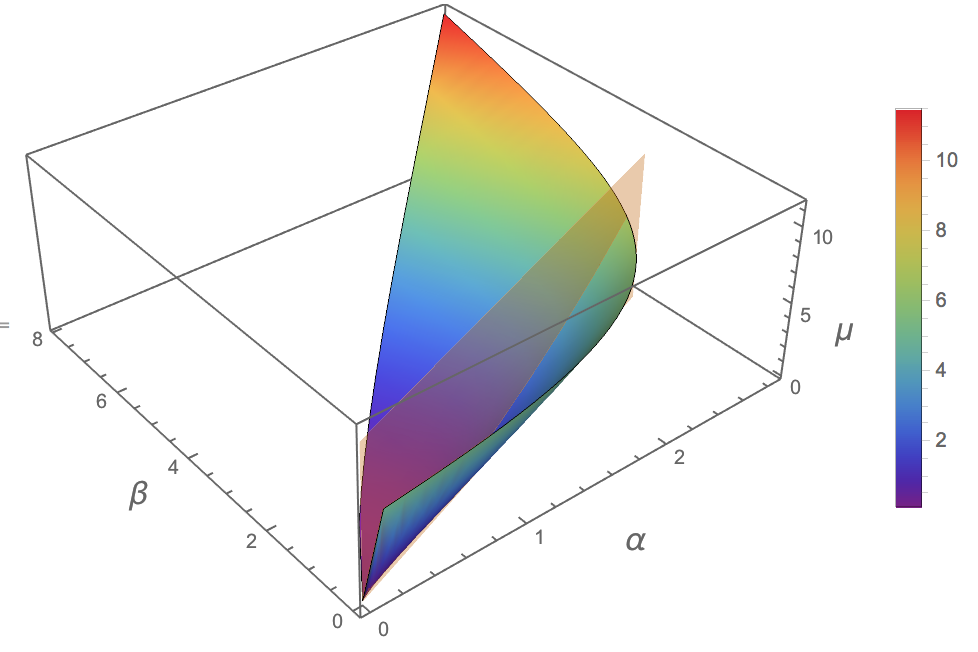}
\caption{\label{fig2} Black hole surface $\mu = \mu(\alpha, \beta)$ and boundary condition surface $\beta = \alpha$.
The color coding corresponds to the magnitude of $\mu$.}
\end{subfigure}
\caption{Black hole line and black hole surface for theories with potential
\eqref{V cosh}.}%
\end{figure}

The AdS background with $\varphi=0$ is maximally supersymmetric. Black hole
solutions dressed with non-trivial scalars can be easily constructed
numerically using a simple shooting method. Doing so for the theories
determined from the potential \eqref{V cosh} we find the black hole line shown
in Fig \ref{fig1}. The function $F(z)$ acquires a minimum at $z_{\mathrm{min}%
}\approx1.532$ at which $F(z_{\mathrm{min}})\approx0.975$. In Fig \ref{fig2}
we display the surface $\mu=\mu(\alpha,\beta)$ alongside with the boundary
condition $\beta=\alpha$. Their intersection defines a uni-parametric family
of black holes as described above. Note that because of the reflection
symmetry $\varphi\rightarrow-\varphi$, the black hole surface satisfies
$\mu(-\alpha,-\beta)=\mu(\alpha,\beta)$ (to ease visualization the region of
negative $\alpha$, $\beta$ is not shown in Fig. \ref{fig2}). Furthermore, our
numerics indicate that black holes with different signs of $\alpha$ and
$\beta$ cannot exist.

\begin{figure}[th]
\label{fig:cs2}\centering
\includegraphics[width=0.6 \linewidth]{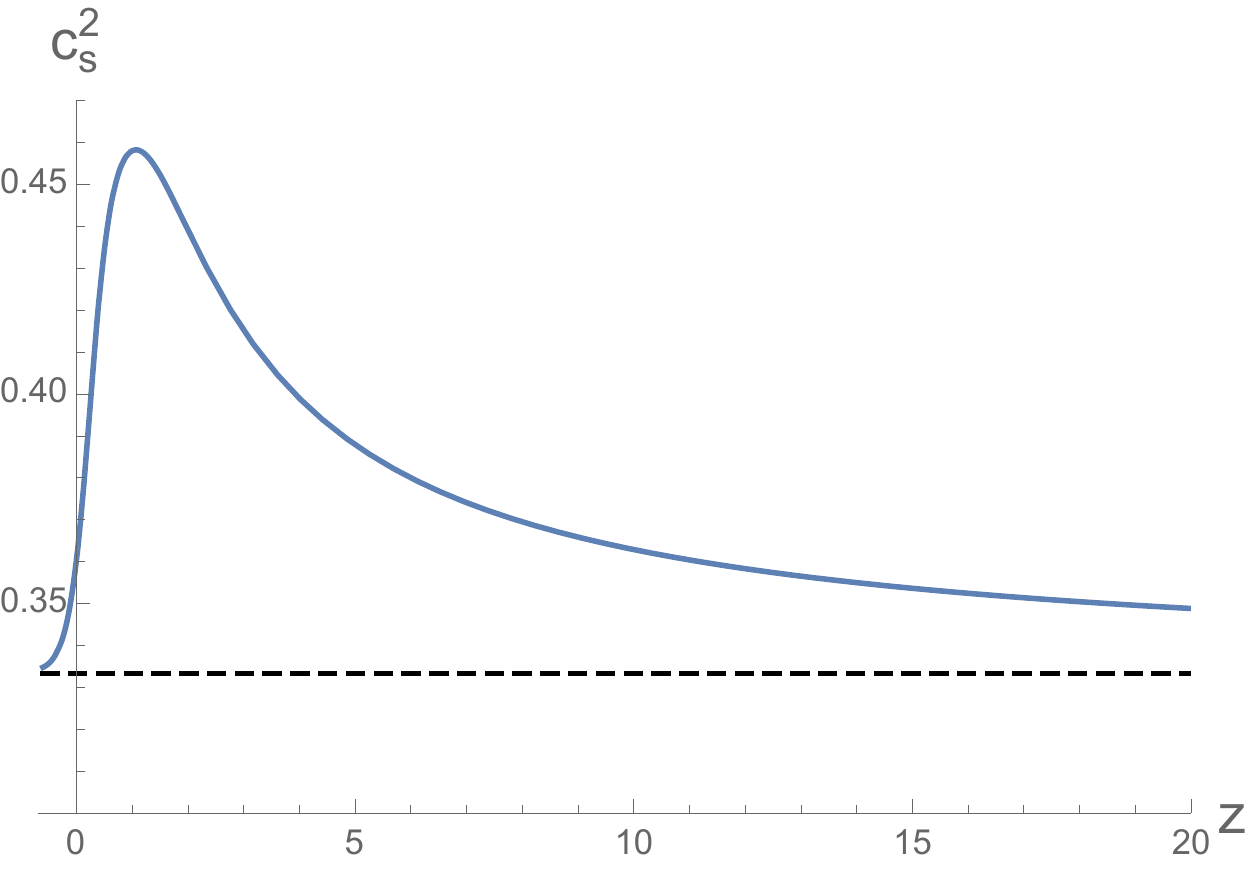}\caption{Speed of sound as a
function of $z$ for the family of boundary conditions $\beta=\lambda\alpha$.
The dashed line depicts the conformal value $c_{s}^{2}=1/3$, which is
approached for large and small $z$.}%
\end{figure}For the linear boundary condition $\beta=\lambda\alpha$, we plot
the speed of sound as a function of $z$ in Fig. 2 --- for this deformation the
energy density is obtained by replacing $W=\frac{\beta^{2}}{2\lambda}$ and
$\alpha=\frac{\beta}{\lambda}$ in (\ref{dens}). This yields $\rho=\frac
{\alpha^{2}l^{3}}{\kappa}\left(  \frac{3}{2}F(z)-\frac{1}{8}\right)  $. This
is indeed an everywhere positive convex function for all the deformations we
are considering. In the plots we have set $z=\frac{1}{2}\ln\left\vert
\alpha\right\vert +1$. So, when $\alpha\gg1$, then $z$ is very large and the
deformed theory explores very high energies, where the conformal value for the
speed of sound is recovered as one would expect. When $\alpha\sim0$ then $z<0$
and the deformation vanishes. The conformal value for the speed of sound is
again recovered in this limit. In the intermediate regime the speed of sound
is larger than the conformal value, so it appears to be a very interacting
state of matter which might be relevant for the description of neutron stars,
where the speed of sound is believed to be bounded only by the speed of light
\cite{Rhoades:1974fn}. We note that the conjectured bound ($c_{s}^{2}\leq1/3$)
for the speed of sound in theories with an holographic dual was only supposed
to hold when there is no chemical potential \cite{Cherman:2009tw}. This bound
was indeed shown to be violated when a chemical potential is introduced in
\cite{Hoyos:2016cob}. In this sense, this is the first example of the
violation of the conjectured bound with an everywhere positive convex energy functional.

\section{The Renormalization group}

The field theory variables $\alpha$ and $\beta$ have an ambiguity when written
in terms of the gravity variable $\varphi\,$, see \eqref{phi1}. This is
parametrized by the constant $r_{0}$ necessary to make the argument of the
logarithm dimensionless. This is interpreted on the field theory side as a
standard one-loop renormalization \cite{Witten:2001ua}. We take the bare
coupling constant $\lambda_{\Lambda}$ and the bare field $\beta_{\Lambda}$ to
be defined at the cutoff scale $\Lambda$ and the renormalized coupling
constant $\lambda_{\nu}$ and renormalized field $\beta_{\nu}$ defined at the
mass scale $\nu$. The independence of the bulk field from the renormalization
procedure yields
\begin{equation}
\varphi l^{-4}=\frac{\beta_{\nu}\left(  \lambda_{\nu}^{-1}\ln\left(
r\nu\right)  +1\right)  }{r^{2}}+O(\frac{\ln(r)^{2}}{r^{3}})=\frac
{\beta_{\Lambda}\left(  \lambda_{\Lambda}^{-1}\ln\left(  r\Lambda\right)
+1\right)  }{r^{2}}+O(\frac{\ln(r)^{2}}{r^{3}}) \label{eq7-2}%
\end{equation}
where the latter equality implies $\beta_{\nu}\lambda_{\nu}^{-1}%
=\beta_{\Lambda}\lambda_{\Lambda}^{-1}$ and $\lambda_{\nu}=\lambda_{\Lambda
}+\ln\left(  \Lambda/\nu\right)  $ or alternatively
\begin{equation}
\lambda_{\nu}^{-1}=\frac{\lambda_{\Lambda}^{-1}}{1+\lambda_{\Lambda}^{-1}%
\ln\left(  \Lambda/\nu\right)  }\text{ ,} \label{lambda}%
\end{equation}
which has the form of a one loop renormalization of a dimensionless coupling
constant \cite{Witten:2001ua}. In the plot of the speed of sound of the
previous section we have set $z=\frac{1}{2}\ln\left\vert \alpha\right\vert
+1$, which means setting the renormalized coupling constant to one and
measuring $\alpha$ in terms of the scale given by $\nu^{2}$:
\begin{equation}
z=\frac{1}{2}\ln\left\vert \frac{\alpha}{\nu^{2}}\right\vert +\lambda_{\nu
}\text{ .} \label{z}%
\end{equation}
The energy scale and the value of the dressed coupling constant are indeed
necessary to compare the theoretical prediction with the experiment. A change
in the values that we picked would imply a simple shift in the graph of figure 2.

\section{Other scaling and spacetime dimensions}

Although we have worked in five dimensions for scalar fields saturating the
Breitenlohner-Freedman bound, our results are easily generalized to any
spacetime dimension, $D$, for any scalar field with masses between this bound
and the unitarity bound. Scalar fields with mass in the BF window
(\ref{window}) fall-off as
\begin{equation}
\varphi\sim\frac{\alpha}{r^{\Delta-}}+\frac{\beta}{r^{\Delta+}}+O(r^{-\Delta
_{-}-1})\text{ ,}%
\end{equation}
where $2<\Delta_{-}<\Delta_{+}=D-1-\Delta_{-}$ and $\frac{\Delta_{+}}%
{\Delta_{-}}$ is not an integer \cite{Henneaux:2006hk}. The energy density is
in these cases \cite{Amsel:2006uf}%

\begin{equation}
\rho\sim\frac{D-2}{2}\mu+\left(  \Delta_{+}-\Delta_{-}\right)  \left[
W(\alpha)-\frac{\Delta_{-}}{D-1}\alpha\beta\right]  \text{ ,}%
\end{equation}
where $\beta=\frac{dW\left(  \alpha\right)  }{d\alpha}$. Using the Smarr
formula of \cite{Liu:2015tqa} and the fact that $\rho-(D-2)p$ vanishes for AdS
invariant boundary conditions yields%

\begin{equation}
p\sim\frac{1}{2}\mu-\left(  \Delta_{+}-\Delta_{-}\right)  \left[
W(\alpha)-\frac{\Delta_{-}}{D-1}\alpha\beta\right]  \text{ .}%
\end{equation}
In these cases the black hole surface takes the form%

\begin{equation}
\alpha=\alpha_{h}\left(  \varphi_{h}\right)  \mathcal{A}^{\frac{\Delta_{-}%
}{D-2}}\text{ ,}\qquad\mu\left(  \varphi_{h},\mathcal{A}\right)  =\mu
_{h}\left(  \varphi_{h}\right)  \mathcal{A}^{\frac{D-1}{D-2}}\text{ ,}%
\qquad\beta=z\left(  \varphi_{h}\right)  \alpha^{\frac{\Delta_{+}}{\Delta-}%
}\text{ .}%
\end{equation}
and using the black hole line $\mu=F(z)\alpha^{\frac{D-1}{\Delta_{-}}}$ and
the boundary condition $z=\omega(\alpha)$, the universal formula for the speed
of sound is%

\begin{equation}
c_{s}^{2}=\frac{1}{D-2}+2\frac{D-1}{D-2}\frac{\omega^{\prime}\alpha
(D-1-2\Delta_{-})\Delta_{-}^{2}}{{(D-2)(D-1)}\left(  {(D-1)F+\omega^{\prime
}\alpha\Delta}_{-}{\dot{F}}\right)  {-2\Delta_{-}^{2}(D-1-2\Delta}_{-}%
{)}\omega^{\prime}{\alpha}}.
\end{equation}
which would allow one to model any given equation of state as for instance the
one given by lattice QCD (for references see the recent review
\cite{Philipsen:2012nu}). We shall provide a detailed derivation in a future work.

\section*{Acknowledgments}

We thank Andrei Starinets for valuable discussions. Research of AA is
supported in part by Fondecyt Grant 1141073 and Newton-Picarte Grants
DPI20140053 and DPI20140115. The work of TA is supported by the European
Research Council under the European Union's Seventh Framework Programme (ERC
Grant agreement 307955). The work of DA is supported by the Fondecyt Grant
1161418 and Newton-Picarte Grant DPI20140115. This work was supported in part
by the Natural Sciences and Engineering Research Council of Canada.


\end{document}